\documentclass[fleqn]{article}
\usepackage{graphicx}
\usepackage{epsfig}
\usepackage{amsmath}
\usepackage{amsthm}
\usepackage{amssymb}
\usepackage{enumerate}
\usepackage{calc}
\usepackage{color}
\usepackage{upgreek}
\newcommand{\beq}{\begin{equation}}
\newcommand{\eeq}{\end{equation}}
\newcommand{\bea}{\begin{eqnarray}}
\newcommand{\eea}{\end{eqnarray}}
\theoremstyle{plain}

\theoremstyle{definition}
\newtheorem{defn}{Definition}
\theoremstyle{remark}

\def\FF{\mathbb{F}}

\usepackage{hyperref}
\hypersetup{
  pdfauthor={Fr\'ed\'eric Holweck},
  pdftitle={Contextuality induced by the skew-embedding of the split Cayley hexagon in the three-qubit Pauli group}
  pdfproducer={LaTeX with hyperref},
  pdffitwindow=false,
  pdfstartview={FitH},
  colorlinks=true,
  linkcolor={blue},
  citecolor={blue},
  filecolor={blue},
  urlcolor={blue}
}
\begin{document}
\Large
\begin{center}
{\bf Three-Qubit-Embedded Split Cayley Hexagon is Contextuality Sensitive}
\end{center}
\large
\vspace*{-.1cm}
\begin{center}
Fr\'ed\'eric Holweck$^{1,2}$, Henri de Boutray$^{3}$ and Metod Saniga$^{4}$ 
\end{center}
\vspace*{-.4cm} \normalsize
\begin{center}

$^{1}$Laboratoire Interdisciplinaire Carnot de Bourgogne, ICB/UTBM, UMR 6303 CNRS, Universit\'e de Bourgogne Franche-Comt\'e, 90010 Belfort Cedex, France

$^{2}$Department of Mathematics and Statistics, Auburn University, Auburn, AL, USA

$^{3}$ ColibriTD, La D\'efense, Paris, France

$^{4}$Astronomical Institute, Slovak Academy of Sciences, SK-05960 Tatransk\'a Lomnica, Slovak Republic

\vspace*{.0cm}

\vspace*{.2cm} (29 January 2022)

\end{center}

\vspace*{-.3cm} \noindent \hrulefill

\vspace*{.1cm} \noindent {\bf Abstract:} 
It is known that there are two non-equivalent embeddings of the split Cayley hexagon of order two into $\mathcal{W}(5,2)$, the binary symplectic polar space of rank three, called classical and skew. Labelling the 63 points of $\mathcal{W}(5,2)$ by the 63 canonical observables of the three-qubit Pauli group subject to the symplectic polarity induced by the (commutation relations between the elements of the) group, the two types of embedding are found to be quantum contextuality sensitive. In particular, we show that the complement of a classically-embedded hexagon is not contextual, whereas that of a skewly-embedded one is. 

\vspace*{.3cm} \noindent
\hspace*{1.95cm} 

\vspace*{-.2cm} \noindent \hrulefill

\section{Introduction}

Quantum contextuality is one of the most counter-intuitive notions in quantum physics that is assumed to be of main importance in quantum information \cite{Howard2014,Bermejo2017,Lillystone2019,Shahandeh2021}. Roughly speaking, quantum contextuality rules out any Hidden Variables Theory unless one assumes that the pre-existing values of a measurement depend on the context, i.e. on the set of mutually commuting measurements in which a given experience takes place.
First proved mathematically by Kochen and Specker \cite{Kochen1975}, several alternative proofs \cite{Lison2014,Cabello2016} and experimental validations \cite{Bartosik2009,Dambrosio2013} have been provided since then. For a recent comprehensive survey about contextuality, see \cite{Budroni2021}.

Among various formulations of quantum contextuality we will deal in this paper with observable-based proofs of quantum contextuality as proposed by Mermin \cite{Mermin1993} and Peres \cite{Peres1990}. In this formulation, see Section \ref{sec:contextuality}, a proof of the Kochen-Specker (KS) Theorem is provided by a configuration of sets of mutually commuting observables, called contexts, such that all the constrains imposed by the configuration on the eigenvalues of those observables cannot be satisfied by a classical function unless that function is context-dependent. We will call such a configuration {\em contextual} when it furnishes a proof of the KS Theorem. These proofs of contextuality can also be tested experimentally \cite{Cabello2010,Dikme2020, Holweck2021}. Interestingly, this type of contextual configurations, which involves $N$-qubit Pauli observables, can be embedded as subgeometries of  the geometric realization of the $N$-qubit Pauli group known as the symplectic polar space of rank $N$ and order two, usually denoted as $\mathcal{W}(2N-1,2)$, see Section \ref{sec:contextuality}. This geometric perspective on $N$-qubit observables was introduced some $15$ years ago \cite{Saniga2007a,Thas2009, Havlicek2009} and led to several surprising connections between physics and geometry \cite{Saniga2004} and also between different branches of modern physics, like, for instance, the so-called black-hole/qubit correspondence \cite{Levay2009,Levay2019}.

In $\mathcal{W}(5,2)$ one can find a number of copies of a very special configuration made of $63$ points and $63$ lines, called the split Cayley hexagon of order two. This configuration is a generalized hexagon is the sense that it is $m$-gon free for $m\leq 5$. The fact that it can be embedded into the three-qubit polar space, $\mathcal{W}(5,2)$, was first employed in \cite{Levay2008} to establish connections between three-qubit observables and black-hole entropy formulas. Later it was also proved that this configuration has, in fact, two distinguished embeddings in $\mathcal{W}(5,2)$, one called classical, the other skew \cite{Coolsaet2010}. In this article one shows that these two distinguished embeddings behave differently in terms of contextuality. More precisely, if one denotes the two embeddings by $\mathcal{H}_C$ and $\mathcal{H}_S$, respectively, then one demonstrates that the configuration $\overline{\mathcal{H}}_C=\mathcal{W}(5,2)\setminus \mathcal{H}_C$, i.e. the configuration made of all the line-contexts that are in $\mathcal{W}(5,2)$ but not in  $ \mathcal{H}_C$, is not contextual, while  $\overline{\mathcal{H}}_S=\mathcal{W}(5,2)\setminus \mathcal{H}_S$ is contextual.

The paper is organized as follows. In Section \ref{sec:contextuality} we recall the basic notions on observable-based proofs of contextuality as well as the notion of symplectic polar spaces where corresponding configurations live. We also explain how we  prove that a given configuration is contextual by solving a particular linear system. In Section \ref{sec:schex} one details the differences between the two distinguished symplectic embeddings of the split Cayley hexagon of order two and recall some known results about these specific configurations. Finally in Section \ref{sec:result} we collect our results and state precisely our hexagon-related contextuality findings. Section \ref{sec:conclusion} concludes this paper with general remarks and perspectives.

\section{Observable-based proofs of contextuality and the symplextic polar spaces over the two-element field}\label{sec:contextuality}
\begin{figure}[!ht]
 \begin{center}
  \includegraphics[width=5cm]{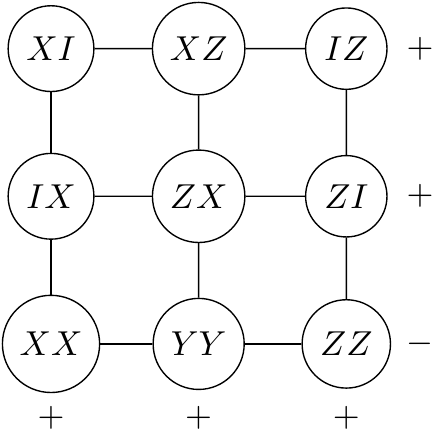}
  \caption{The Peres-Mermin magic square: All two-qubit observables on the same row/column mutually commute. The product of observables in each row/column is equal to $\pm I_4$ as indicated by the signs. The measurement of each two-qubit observable yields $\pm 1$. One observes that there is no deterministic classical function that would predict the outcomes for each node and satisfy all the signs constraints.}\label{fig:magicsquare}
 \end{center}
\end{figure}

Let us recall the principle of the Mermin and Peres \cite{Mermin1993,Peres1990} operator-based proofs of quantum contextuality. Consider the configuration of two-qubit observables depicted in Figure \ref{fig:magicsquare} and known as a Peres-Mermin magic square.

\begin{figure}[!ht]
 \begin{center}
  \includegraphics[width=6cm]{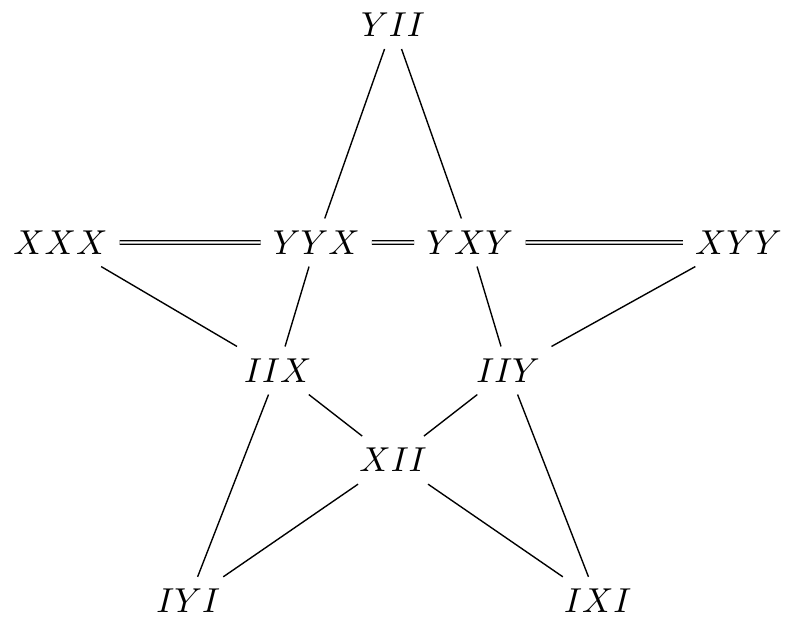}
  \caption{The Mermin pentagram \cite{Mermin1993}. Each node is labelled by a three-qubit Pauli observable. The lines of the pentagram are sets of mutually commuting observables whose product is $+I_8$ (each thin line) or $-I_8$ (the double line). The fact that there is only one (odd number) negative line allows us to use the same argument as for the Peres-Mermin magic square. This configuration therefore also furnishes a proof of the KS Theorem.}\label{fig:pentagram}
 \end{center}
\end{figure}

 Each node of the square is labelled by a two-qubit observable with a shorthanded notation $AB\equiv A\otimes B$. Here $A,B\in \{I,X,Y,Z\}$ with $X,Y,Z$ being the usual Pauli matrices and $I$ the unit matrix. In each row/column of the grid, two-qubit observables mutually commute and form what we call a context, i.\,e. a set of compatible measurements whose product is, up to a sign, the identity operator. In the Peres-Mermin square the product of the observables on each context is $\pm I_4$ as indicated by the signs in Figure \ref{fig:magicsquare}. In each context, the product of the measurements, which are eigenvalues of the observables, should be an eigenvalue of the product of the observables. The eigenvalues of each node are $\pm 1$ and the constraint on each context is $\pm 1$ according to the signs prescription on rows/columns. Because there is an odd number of negative contexts in the square, it is clear that there exists no classical function that can assign predefinite values to each node and satisfy all the constrains imposed by the signs of the contexts (rows/columns). If such a classical function existed, it would necessarily be context-dependent, i.\,e. the predefined value assigned to a given node would depend on the considered context. This elementary configuration of observables furnishes a proof of the KS Theorem by showing that there is no Non-Contextual Hidden Variable (NCHV) theory that can reproduce the outcomes of quantum physics. Another elementary proof based on three-qubit observables was proposed by Mermin in \cite{Mermin1993} and is reproduced in Figure \ref{fig:pentagram}.

In \cite{Holweck2017} it was proved that Peres-Mermin magic squares and Mermin pentagrams are the smallest possible proofs of the KS Theorem that can be achieved in terms of the number of observables and contexts: $9$ observables and $6$ contexts for the Peres-Mermin magic square and $10$ observables and $5$ contexts for the pentagram.

If one considers two-qubit Pauli observables up to a global phase ($\pm 1$, $\pm i$), there are $10$ copies of the magic square. One explains now how one can naturally embed these $10$ grids into the symplectic polar space of rank $N=2$ over the two-elements field $\mathbb{F}_2=\{0,1\}$. In full generality let us consider the group of $N$-qubit Pauli observables, i.e.
\begin{equation}\label{eq:pn}
 \mathcal{P}_N=\{s A_1A_2 \dots A_N, s\in \{\pm 1,\pm i\}, A_i \in \{I,X,Y,Z\}\},
\end{equation}
employing again the shorthand notation $A_1A_2\dots A_N\equiv A_1\otimes A_2\otimes \dots \otimes A_N$. 
The center of $\mathcal{P}_N$ is $\mathcal{C}_N=\{\pm I_{2^N}, \pm i I_{2^N}\}$. The abelian group $\mathcal{P}_N/\mathcal{C}_N$ is isomorphic to the $2N$-dimensional vector space $V_N=\mathbb{F}_2 ^{2N}$ as we now explain.

Up to a phase, each observable $A_i$ can be represented by a doublet $(\mu_i,\nu_i)\in \mathbb{F}_2^2$ as follows
\begin{equation}
 A_i=Z^{\mu_i}\cdot X^{\nu_i} \text{ where } \cdot \text{ denotes the ordinary product of matrices};
\end{equation}
more precisely, we have the following relations:
\begin{equation}
 I\leftrightarrow (0,0), X\leftrightarrow(0,1), Z\leftrightarrow(1,0),iY\leftrightarrow(1,1).
\end{equation}
Hence, to a given  class $\overline{\mathcal{O}}=\{A_1A_2\dots A_N, -A_1A_2\dots A_N, iA_1A_2\dots A_N, -iA_1 A_2 \dots A_N\}$ in $\mathcal{P}_N/\mathcal{C}_N$ one can associate a unique $2N$-plet $(\mu_1,\nu_1,\dots,\mu_N,\nu_N) \in \mathbb{F}_2^{2N}$. Moreover, the multiplicative structure of  $\mathcal{P}_N/\mathcal{C}_N$ is mapped to the additive structure of $V_N$. Considering now the projective space associated with $V_N$, $PG(2N-1,2)$, one obtains a map between non-trivial $N$-qubit observables, up to a global phase, and points of $PG(2N-1,2)$
\begin{equation}
 \pi: \left\{\begin{array}{ccc}
   (\mathcal{P}_N/\mathcal{C}_N)\setminus I_{2^N} & \to & PG(2N-1,2),\\
   \mathcal{O} & \mapsto & [\mu_1:\nu_1:\mu_2:\mu_2:\dots:\mu_N:\nu_N].
  \end{array}\right.
\end{equation}

This mapping represents (classes of) observables as points in $PG(2N-1,2)$, but tells us nothing about the commutation relations in $\mathcal{P}_N$. In order to recover this information, one introduces the following symplectic form on 
$PG(2N-1,2)$: 
\begin{equation}\label{eq:symplectic}
  \langle p,q\rangle=\sum_{i=1}^N p_iq_{N+i}+p_{N+i}q_i,
\end{equation}
where $p=[p_1:\dots:p_{2N}]$ and $q=[q_1:\dots:q_{2N}]\in PG(2N-1,2)$.
Then, if one considers two non-trivial classes $\overline{\mathcal{O}}_p$ and $\overline{\mathcal{O}_q}$  of $N$-qubit Pauli observables in $\mathcal{P}_N/\mathcal{C}_N$ such that $\pi(\overline{\mathcal{O}}_p)=p$ and $\pi(\overline{\mathcal{O}}_q)=q$, then a straightforward calculation shows that 
\begin{equation}
 \mathcal{O}_p \text{ and } \mathcal{O}_q \text{ commute }\Leftrightarrow \langle p,q\rangle=0.
\end{equation}
Employing the above-defined symplectic form leads to the definition of $\mathcal{W}(2N-1,2)$, the symplectic polar space of rank $N$ and order $2$.
\begin{defn}
 The space of totally isotropic subspaces\footnote{A linear space is said to be totally isotropic if and only if the symplectic form vanishes identically on the space.} of $PG(2N-1,2)$ endowed with a nondegenerate symplectic form $\langle,\rangle$ is called the symplectic polar space of rank $N$ and order $2$, $\mathcal{W}(2N-1,2)$.
\end{defn}

Note that because of the symplectic form, the group $\text{Sp}(2N,2)$ of symplectic matrices acts transitively on $\mathcal{W}(2N-1,2)$. This group is spanned by the transvections $T_p$ for all $p\in \mathcal{W}(2N-1,2)$:

\begin{equation}
 T_p:\left\{\begin{array}{ccc}
\mathcal{W}(2N-1,2) & \to & \mathcal{W}(2N-1,2), \\
 q  & \mapsto & q+\langle p,q\rangle p.
   \end{array}\right.
\end{equation}
If we consider the action of transvections in terms of the labelling of $\mathcal{W}(2N-1,2)$ by $N$-qubit Pauli observables, then for $p=\pi(\overline{\mathcal{O}}_p)$ and $q=\pi(\overline{\mathcal{O}}_q)$ we have

\begin{equation}
  T_{\mathcal{O}_p}(\overline{\mathcal{O}}_q)=
  \left\{\begin{array}{ccc}
    \overline{\mathcal{O}}_q & \text{ if and only if } & \overline{\mathcal{O}}_q \text{ and } \overline{\mathcal{O}}_q \text{ commute},\\
    \overline{\mathcal{O}_q\cdot\mathcal{O}_p} & \text{ if and only if } & \overline{\mathcal{O}}_q \text{ and } \overline{\mathcal{O}}_q \text{ anticommute}.
  \end{array}\right.
\end{equation}
The points of $\mathcal{W}(2N-1,2)$ are $N$-qubit observables, lines correspond to triplet of mutually commuting elements whose product is $\pm I_{N}$ and form contexts. Planes and higer dimensional linear subspaces of $\mathcal{W}(2N-1,2)$ can also be of some use to generate contexts. Let us illustrate this for $N=2$ and $N=3$ \cite{Saniga2007,Levay2017}.

\subsection{
\texorpdfstring{$N=2$, the Doily $\mathcal{W}(3,2)$}{N=2, the Doily W(3,2)}
} 

\begin{figure}[!ht]
 \begin{center}
  \includegraphics[width=5cm]{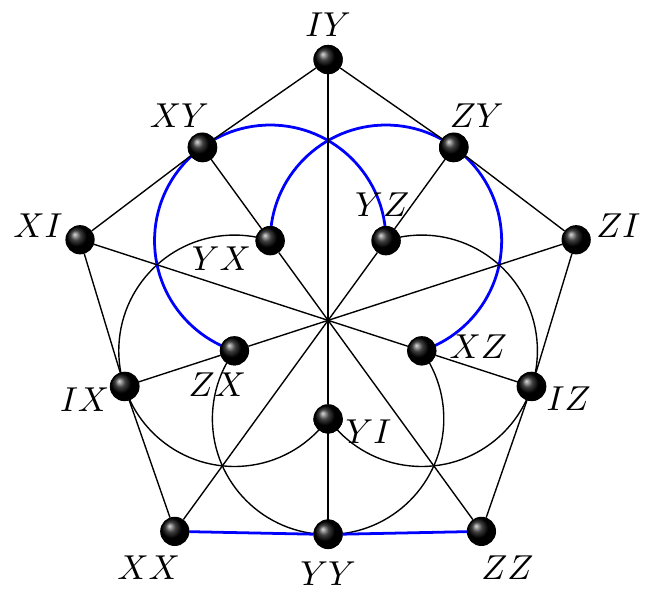}
  \caption{The symplectic polar space $\mathcal{W}(3,2)$ labelled by two-qubit Pauli observables. Each line of the configuration forms a context. The three negative contexts are indicated in blue.}\label{fig:doily}
 \end{center}
\end{figure}

The symplectic polar space of rank $2$ and order $2$, known as the doily, encapsulates the commutation relations within the two-qubit Pauli group. This space features $15$ points and $15$ lines and form a $15_3$ configuration, i.e. a point-line incidence structure with $3$ points per line and $3$ lines per point. Out of $245,342$ non isomorphic $15_3$-configurations, the doily is the only one to be {\it triangle free}. Using the canonical representatives ($s=1$ in (\ref{eq:pn})), one gets a labelling of the doily by two-qubit Pauli observables as given in Figure \ref{fig:doily}.

\begin{figure}[!ht]
\begin{center}
 \includegraphics[width=5cm]{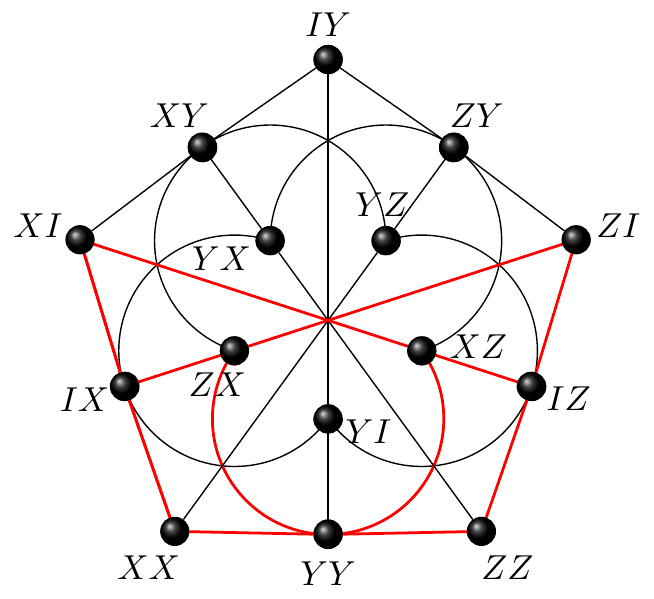}
 \caption{A Peres-Mermin square as a subgeometry of the doily. The square is the same as the one introduced in Figure \ref{fig:magicsquare}. It is a geometric hyperplane in the sense that a line of $\mathcal{W}(3,2)$ is either contained in it (six such lines depicted in red) or has one point common with it (each of the remaining nine lines).}\label{fig:mermininW}
 \end{center}
\end{figure}

Peres-Mermin magic squares are subgeometries of $\mathcal{W}(3,2)$. More precisely they are geometric hyperplanes of $\mathcal{W}(3,2)$, i.e. sets of points such that a line of $\mathcal{W}(3,2)$ is either fully contained in such set or shares with it a single point. An example of such a set is given in Figure \ref{fig:mermininW}. Starting with this canonical labeling one can form $9$ more copies of Peres-Mermin square by acting by transvections. Each of them have an odd number, namely $1$ or $3$, of negative contexts and is thus contextual.

To conclude this subsection it is worth mentioning that the space $\mathcal{W}(3,2)$ of its own is a contextual configuration, as first proved in \cite{Cabello2010}. In fact, Cabello shows that the configuration corresponding to points and lines of $\mathcal{W}(2N-1,2)$ is contextual for any $N \geq 2$.

\subsection{
\texorpdfstring{$N=3$, the symplectic polar space $\mathcal{W}(5,2)$}{N=3, the symplectic polar space W(5,2)}
}

\begin{figure}[!ht]
 \begin{center}
  \includegraphics[width=5cm]{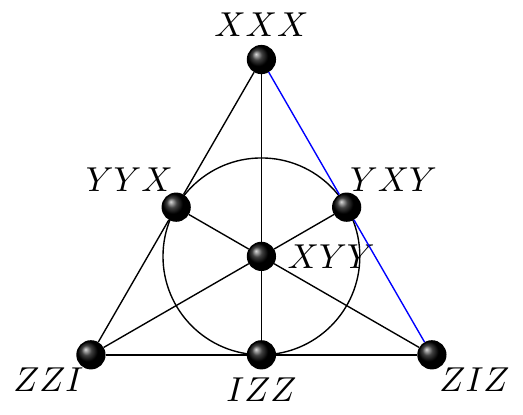}
  \caption{An example of the (negative) Fano plane in $\mathcal{W}(5,2)$; the blue line is also negative.}\label{fig:fano}
 \end{center}
\end{figure}

The symplectic polar space of rank $3$, $\mathcal{W}(5,2)$, comprises $63$ points, $315$ lines and $135$ Fano planes. An interesting study of geometric hyperplanes of $\mathcal{W}(5,2)$ and their physical interpretations in terms of representation theory and invariants can be found in \cite{Levay2017}. Fano planes of $\mathcal{W}(5,2)$ are totally isotropic $2$-dimensional spaces over $\mathbb{F}_2$. An example of such a plane is given in Figure \ref{fig:fano}.
The product of the seven observables in a Fano plane yields $\pm I_{8}$. If one removes a line, one gets a new type of context -- an affine plane of order two -- made of four observables whose product is $\pm I_8$. For instance, in the negative Fano plane of Figure \ref{fig:fano}, removing the line $ZZI-IZZ-ZIZ$, which is positive, leads to the context $XXX-YYX-YXY-XYY$ that is exactly the negative line of the Mermin pentagram of Figure \ref{fig:pentagram}. Using three-qubit Pauli operator contexts of this type one can create altogether $12,096$ distinct Mermin's pentagrams in $\mathcal{W}(5,2)$, as it was first shown by computer calculations \cite{Planat2013} and later proved also rigorously by sole geometric arguments \cite{Levay2013}.
In \cite{deBoutray2021} it was demonstrated that geometric hyperplanes of $\mathcal{W}(5,2)$ defined by quadratic equations (hyperbolic or elliptic) are also contextual configurations; contextual inequalities for the  hyperbolic case were also tested on a quantum computer \cite{Holweck2021}.

In Section \ref{sec:schex} we will introduce another prominent configuration, namely the split Cayley hexagon of order two, which lives in $\mathcal{W}(5,2)$ in two distinct modes featuring {\it different} contextuality behavior. 

\subsection{A simple method for checking contextuality of a configuration}

In order to prove that a particular point-line configuration of observables, i.e. a set of observables $\{\mathcal{O}_j\}_{j\in J}$ grouped into contexts $\{\mathcal{C}_i\}_{i\in I}$, provides a KS proof, we reformulate the question as a linear problem \cite{deBoutray2021}. Indeed, the existence of an NCHV theory that accommodates all contexts $\mathcal{C}_i$ of a configuration means that there exists a classical function, $f$, such that $f(\mathcal{O}_{j})=\pm 1$ for all observables and for all contexts $\mathcal{C}_i$ of the configuration. This means that $f$ satisfies the following system of equations:
\begin{equation}\label{eq:sysclassic}
\Pi_{\mathcal{O}_j\in \mathcal{C}_i}f(\mathcal{O}_j)=\text{sign}(\mathcal{C}_i),
\forall i\in I,\end{equation}
where $\text{sign}(\mathcal{C}_i)$ is the sign of the context $\mathcal{C}_i$.

Consider now the incidence matrix $A$ of the configuration, i.\,e. an $I\times J$ matrix where $a_{ij}=1$ if and only if $\mathcal{O}_j\in \mathcal{C}_i$ and $a_{ij}=0$ otherwise. Let $b\in \FF_2^I$ such that $b_i=0$ if and only if $\text{sign}(\mathcal{C}_i)=1$ and $b_i=1$ if and only if $\text{sign}(\mathcal{C}_i)=-1$. Then the existence of a classical function $f$ that satifies the constrains of Eq. (\ref{eq:sysclassic}) boils down to finding a vector $x\in \FF_2^J$ which is the solution of the following linear system
\begin{equation}
Ax=b.
\end{equation}
For instance, in the case of the Peres-Mermin configuration of Figure \ref{fig:magicsquare} the linear system to be solved over the two-element field is
\begin{equation}
\begin{pmatrix}
  1 & 1 & 1 & 0 & 0 & 0 & 0 & 0 & 0\\
  0 & 0 & 0 & 1 & 1 & 1 & 0 & 0 & 0\\
  0 & 0 & 0 & 0 & 0 & 0 & 1 & 1 & 1\\
  1 & 0 & 0 & 1 & 0 & 0 & 1 & 0 & 0\\
  0 & 1 & 0 & 0 & 1 & 0 & 0 & 1 & 0\\
  0 & 0 & 1 & 0 & 0 & 1 & 0 & 0 & 1\\
\end{pmatrix}\cdot \begin{pmatrix}
  x_1\\
  x_2\\
  x_3\\
  x_4\\
  x_5\\
  x_6\\
  x_7\\
  x_8\\
  x_9
\end{pmatrix} =\begin{pmatrix}
  0\\
  0\\
  0\\
  0\\
  0\\
  1
\end{pmatrix}.
\end{equation}
One readily sees that this system has no solution as the sum of the first five rows of $A$ equals the last row, while the sum of the first five coordinates of $b$ does not equal the last coordinate. The lack of solution means that the corresponding configuration is contextual, i.\,e. it furnishes a proof of the KS Theorem.
This linear formulation of the problem allows us to get a relatively fast computer-based check  of the contextual nature of larger sets of contexts and observables. Note that the incidence matrix $A$ encodes the geometry of the configuration, while the vector $b$ encodes the signs of the contexts; the latter, of course, depends on the choice of labeling of the points of the symplectic polar space by $N$-qubit observables. 

\section{The split Cayley hexagon of order two and its two non-equivalent symplectic embeddings}\label{sec:schex}

A generalized $n$-gon $\mathcal{G}$ of order $(k,l)$ is an point-line incidence structure such that every line contains $k+1$ points, every point is contained in $l+1$ lines and $\mathcal{G}$ does not contain any ordinary $m$-gons  for $2\leq m<n$, but two points, two lines or a point and a line are always contained in an $n$-gon \cite{mald}. When $k=l$ one says that the order of $\mathcal{G}$ is $k$.
 The Fano plane, Figure \ref{fig:fano}, is the unique example of the generalized triangle of order two and the doily $\mathcal{W}(3,2)$, Figure \ref{fig:doily}, is the unique (self-dual) generalized quadrangle of order two. There is no generalized $5$-gon of order two, but there exist two generalized hexagons of order two, the split Cayley hexagon and its dual \cite{mald}.
We will focus now on the split Cayley hexagon, $\mathcal{H}$, which also lives in $\mathcal{W}(5,2)$; this means that one can label the $63$ points of $\mathcal{H}$ by the $63$ non trivial three-qubit observables in such a way that the lines of the hexagon are lines of $\mathcal{W}(5,2)$, i.\,e. sets of mutually commuting three-qubit observables.

\begin{figure}[!ht]
 \begin{center}
  \includegraphics[width=10cm]{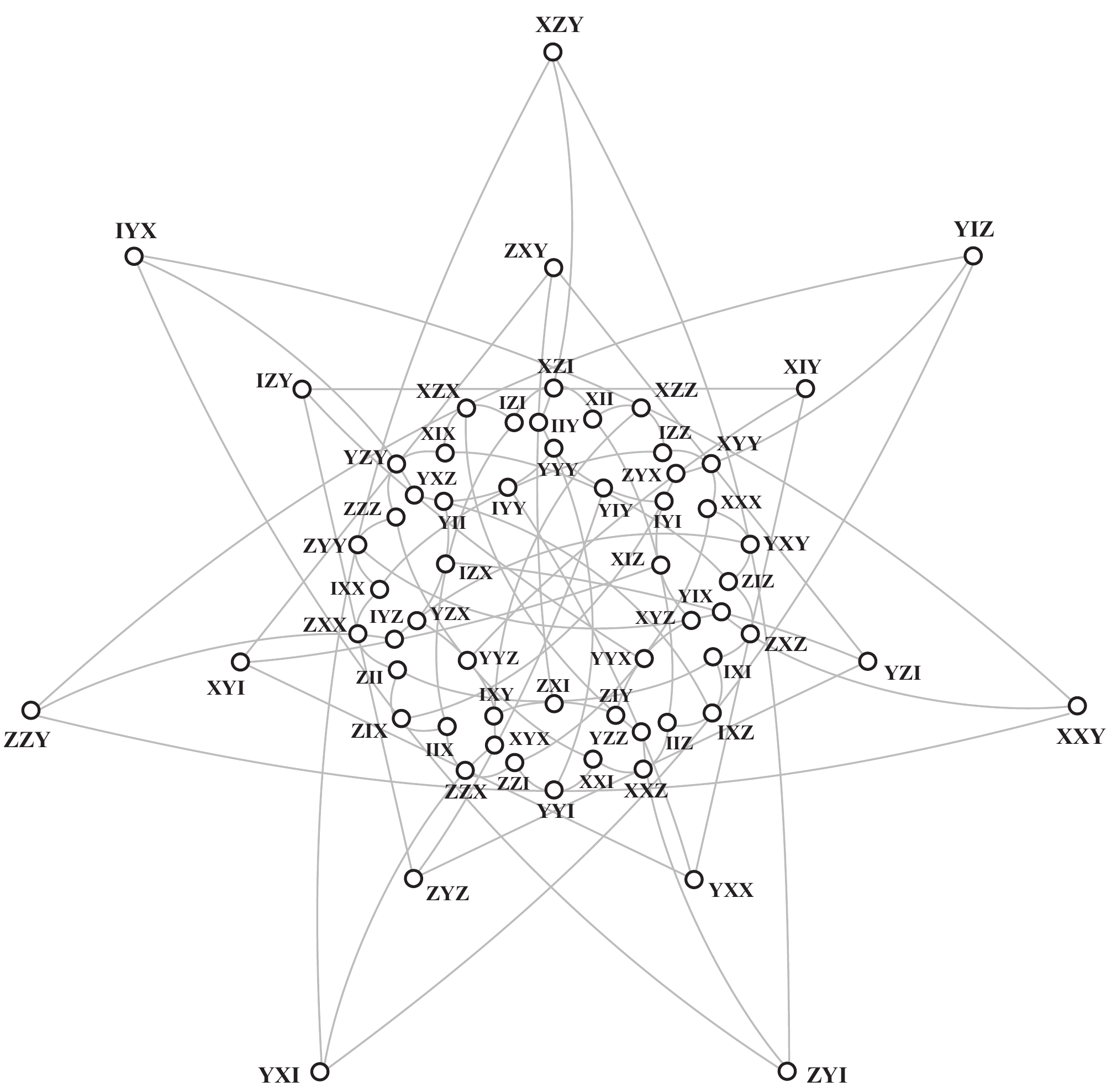}
  \caption{Three-qubit {\em classical} embedding of the split Cayley hexagon of order two as described in \cite{Levay2008}. The
  illustration of points and lines of the hexagon follows closely that of Ref. \cite{schr}}\label{fig:schex}.
 \end{center}
\end{figure}

As already stressed, there are two unequivalent embeddings of $\mathcal{H}$ in $\mathcal{W}(5,2)$.
The first one, called classical,  was explicitly  worked out in \cite{Levay2008} and in the context of quantum information further discussed, for instance, in \cite{Levay2009,Planat2013, Saniga2012}; an example of such embedding is portrayed in  Figure \ref{fig:schex}.

In $\mathcal{W}(5,2)$ there are $120$ classically-embedded copies of $\mathcal{H}$; each of them can be obtained from the copy shown in Figure \ref{fig:schex} by the action of the symplectic group $\text{Sp}(6,2)$. The labelling of Figure \ref{fig:schex} was obtained from the observation that the split Cayley hexagon contains a copy of the Heawood graph, the incidence graph of the Fano plane, and following the procedure outlined in \cite{Polster2001}.
 
Another description of this embedding is the following one \cite{mald}. In $PG(6,2)$, consider the following parabolic quadric $\mathcal{Q}$ 
 \begin{equation}\label{eq:quadric}
  x_1x_4+x_2x_5+x_3x_6-x_7^2=0.
 \end{equation}
 This quadric contains $63$ points. Consider now the lines of $\mathcal{Q}$ that satisfy the following Pl\"ucker equations:
 \begin{equation}\label{eq:plucker}
  \begin{array}{cccc}
   p_{62}=p_{17}, & p_{13}=p_{72}, & p_{24}=p_{37}, & p_{35}=p_{74},\\
   p_{46}=p_{57}, & p_{51}=p_{76}, & p_{14}+p_{25}+p_{36}=0, &
  \end{array}
 \end{equation}
where $p_{ij}=x_ix_j-x_jx_i$. There are $63$ such lines and the obtained $63_3$ configuration is isomorphic to $\mathcal{H}$. This configuration can be bijectively projected into $PG(5,2)$ and $\mathcal{W}(5,2)$. Indeed the projection
  \begin{equation}
   [x_1:\dots:x_7]\mapsto [x_1:\dots:x_6]
  \end{equation}
is a bijection over $\FF_2$ as $x_7=x_1x_2+x_3x_4+x_5x_6$ and the lines of $\mathcal{Q}$ that satisfy the Pl\"ucker relations (\ref{eq:plucker}) are mapped to totally isotropic lines of $PG(5,2)$, i.\,e. the lines of $\mathcal{W}(5,2)$ for the sympletic form (\ref{eq:symplectic}). This embedding will be denoted by $\mathcal{H}_C$.

\begin{figure}[!ht]
 \begin{center}
  \includegraphics[width=10cm]{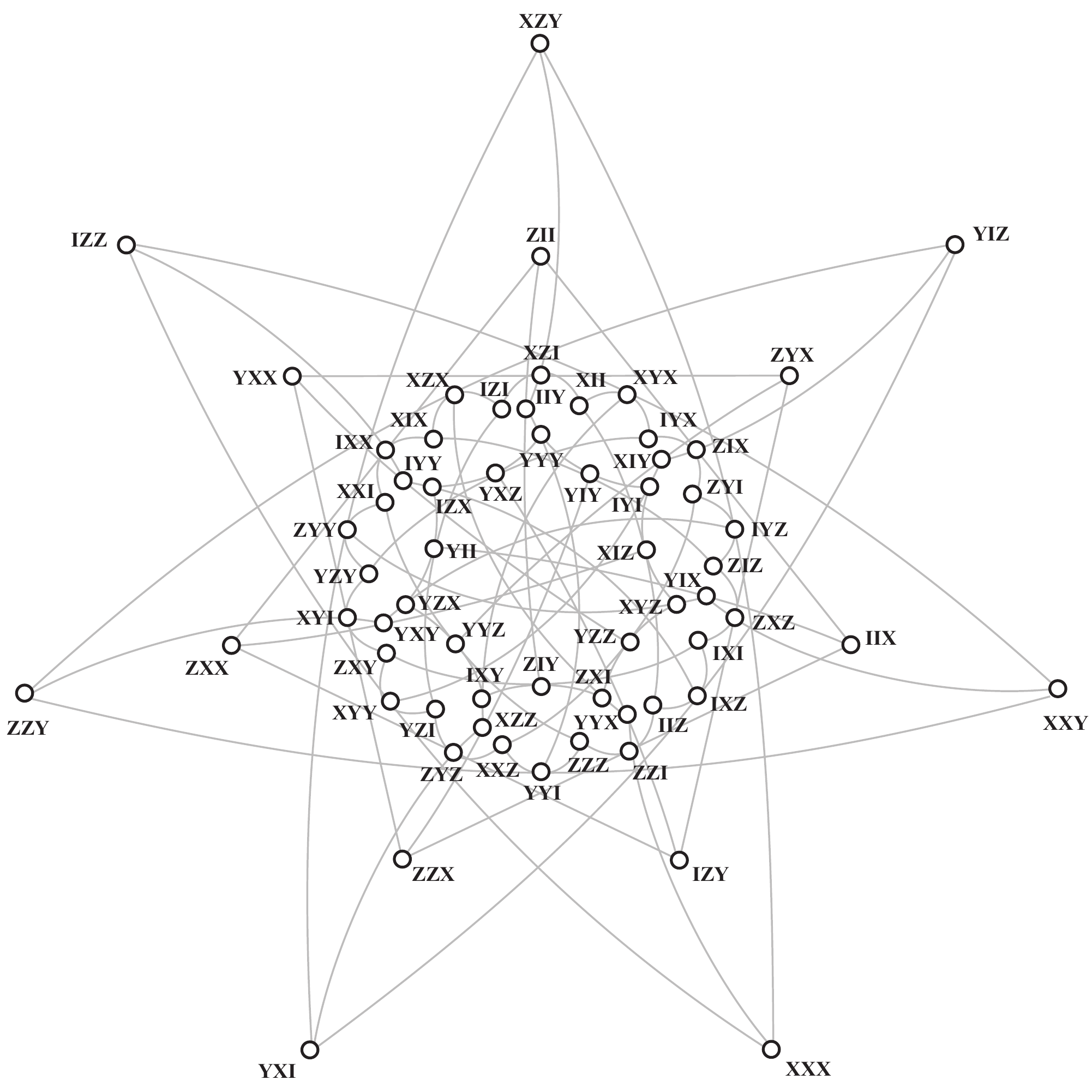}
  \caption{A copy of the split Cayley hexagon of order two that is skew-embedded into $\mathcal{W}(5,2)$.}\label{fig:skewschex}
 \end{center}
 \end{figure}

Interestingly, there exists another, non-equivalent embedding of $\mathcal{H}$ into $\mathcal{W}(5,2)$, as discovered by Coolsaet \cite{Coolsaet2010}. Coolsaet considers, in $PG(6,2)$, the following coordinate map
\begin{equation}\label{eq:map}
 \epsilon:[x_1:\dots:x_7]\mapsto [x_1+x_6+f_5(x):x_2+x_3+f_4(x):x_3:x_4:x_5:x_6:x_7]
\end{equation}
with $f_4(x)=x_3x_5+x_7x_4$ and $f_5(x)=x_4x_6+x_7x_5$, which indeed establishes a different type of embedding of $\mathcal{H}$
 into $\mathcal{Q}$, called skew; its projection to $\mathcal{W}(5,2)$ will be denoted by $\mathcal{H}_S$. There are altogether $7560$ copies of $\mathcal{H}_S$ in $\mathcal{W}(5,2)$; one of them is depicted in Figure \ref{fig:skewschex}.

We shall conclude this section by a brief description of the main difference between the two embeddings. In $\mathcal{H}_C$, all the three lines passing through any of its points are coplanar, i.\,e. lie in the same plane of $\mathcal{W}(5,2)$. This is, however, not the case for
$\mathcal{H}_S$, where only 15 points exhibit this property; for each of the remaining 48 points, only two of the three lines 
are coplanar. To illustrate this, let us take the point $XXX$ of the copy shown in Figure \ref{fig:skewschex}. The three lines passing through it are $\{XXX, XYY, IZZ\}$, $\{XXX, ZZI, YYX\}$ and $\{XXX, IYZ, XZY\}$; clearly, only the first two lie in the same plane. Next, take any observable and all the 30 other observables that commute with it. These 31 observables will form
a geometric hyperplane in both $\mathcal{H}_C$ and $\mathcal{H}_S$. But the two embeddings differ in that while this hyperplane
is of the same type for each point of $\mathcal{H}_C$, an $\mathcal{H}_S$ features two different kinds of them.

\section{
\texorpdfstring{$\overline{\mathcal{H}_S}$ is contextual, whereas $\overline{
 \mathcal{H}}_C$ is not!}{The complement of HS is contextual, whereas the
 complement of HC is not!}
}\label{sec:result}

At this point an interesting question arises: Can the two embeddings be ascribed some physical distinction? To answer this question (in affirmative), we ran several different experiments on $\mathcal{W}(5,2), \mathcal{H}_C's, \mathcal{H}_S's$ and their complements. By the complement, $\overline{\mathcal{G}}$, of a configuration $\mathcal{G}\subset \mathcal{W}(2N-1,2)$ we mean the set of all line-contexts of $\mathcal{W}(2N-1,2)$ that are not in $\mathcal{G}$. We tested by computer  the contextual nature of all $\mathcal{H}_C, \mathcal{H}_S,\overline{\mathcal{H}}_C, \overline{\mathcal{H}}_S$ of $\mathcal{W}(5,2)$ and the results of our analysis are summarized in Table \ref{tab:context}.
 \begin{table}[!ht]
 \begin{center}
  \begin{tabular}{|c|c|c|}
  \hline
   Geometry &  Contextual & \# of Copies \\
   \hline
   $\mathcal{H}_C$ & No & $120$\\
   $\mathcal{H}_S$ & No &$7560$\\
   $\overline{\mathcal{H}}_C$ & No & $120$\\
   $\overline{\mathcal{H}}_S$ & Yes & $7560$\\
   \hline
  \end{tabular}
  \caption{Results on the contextuality properties of differently-embedded hexagons and their complements. The difference between the two embeddings reveals itself in terms of contextuality when one considers the complement of the configuration.}\label{tab:context}
\end{center}
 \end{table}

Our calculations were performed with the software Magma and the computation
resources of the supercomputer of the Mésocentre de calcul de Franche-Comté.
All our codes and results are available on the QuantCert GitHub page\footnote{
URL: \url{https://quantcert.github.io/Magma-contextuality/}}. To implement the
incidence structures of $\mathcal{H}_C$ and $\mathcal{H}_S$ we first encoded the
description in coordinates of the two embeddings provided by Coolsaet 
\cite{Coolsaet2010} (Eq. (\ref{eq:quadric}), Eq. (\ref{eq:plucker}) and Eq.
(\ref{eq:map})). Then we created all $120$ copies of $\mathcal{H}_C$  and all 
$7560$ copies of $\mathcal{H}_S$  by repeatedly acting by transvections. Their
complements were easily obtained from the implementation of $\mathcal{W}(5,2)$. 
Once all incidence structures were created we checked their contextuality by
using the function \texttt{IsConsistent} that establishes if a given
configuration is contextual by employing the procedure described in Section 
\ref{sec:contextuality}.

The \texttt{IsConsistent} Magma intrinsic (compiled function) checks if the rank
of the augmented matrix $(A|b)$ is greater that the one of the coefficient
matrix $A$. This operation is in $O(|\mathcal{O}|+|\mathcal{C}|)$ operations;
in other words, its execution lasts a time proportional to the number of
observables and contexts of the geometry. Given the fact that this precise
system had $315-63=252$ contexts, the duration of this operation is quite
negligible, even repeated several thousands of times. Building the 7560 copies
of $\mathcal{H}_S$ was more intensive though: to reach all copies we started
with a single copy of $\mathcal{H}_S$ and ran every possible combination of $4$
transvections, including the identity, for a total of $64^4=2^24\approx17E6$
total operations. This computation lasted around 11 hours (10.87 to be more
precise), which speaks for the efficiency of Magma.

\section{Conclusion}\label{sec:conclusion}

We have found a very important physical property regarding  the two non-equivalent embeddings of a very special subconfiguration of the symplectic polar space $\mathcal{W}(5,2)$, the split Cayley hexagon of order two. Using the interpretation in terms of three-qubit Pauli observables we showed that the complement of any skew-embedded hexagon is a contextual configuration, i.\,e. provides a proof of the Kochen-Specker Theorem. We tested our findings on all possible embeddings of the split Cayley hexagons and found out that only skew embeddings enjoy this property. To extend this work one may try to measure the degree of contextuality of $\overline{\mathcal{H}}_S$. The degree of contextuality indicates how far is a given contextual configuration to be satisfiable, i.e. non-contextual, if one changes the constrains imposed by the vector $b$. In other words,  if $|J|$ is the number of contexts and $P$ the maximum number of constraints that can be satisfied, then the degree of contextuality is $d=|J|-P$. For a non-contextual configuration $d=0$. The calculation of $d$ boils down to finding the Hamming distance between $b$ and the image of $A$, the incidence matrix of the configuration. However, in the case of the split Cayley hexagon a brute force calculation to compute $d$ is out of reach of the supercomputer resources we have currently at our disposal. It, therefore, requires a deeper understanding of the geometry to reduce the calculation cost. The degree of contextuality $d$ is also necessary for calculating the classical bound of the contextual inequalities given in \cite{Cabello2010} as well as for a possible testing of these inequalities on a quantum computer \cite{Holweck2021}.

\section{Acknowledgements}

This work was supported by the Conseil R\'egional de Bourgogne Franche-Comt\'e, mobility grant {\bf GeoQuant}, the Thomas Jefferson Fundation and by the Slovak VEGA Grant Agency, Project $\#$ 2/0004/20. Part of this work was done during a visit of HdB and a one-year stay of FH at the Mathematics and Statistics Department of Auburn University. Both authors wish to acknowledge the people of the department that made our work here possible, in particular Dr. Luke Oeding for his precious help and friendship as well as Drs. Gaetan Bakalli and Mucyo Karemera for their hospitality and enthusiastic discussions about mathematics and life.

\end{document}